\documentclass[utf8]{frontiersFPHY} 

\usepackage{url,hyperref,microtype,subcaption}
\usepackage[onehalfspacing]{setspace}

\def\keyFont{\fontsize{8}{11}\helveticabold }
\def\firstAuthorLast{De Zotti {et~al.}} 
\def\Authors{G. De Zotti\,$^{1,*}$, M. Bonato\,$^{2,3,1}$, M. Negrello\,$^{4}$, T. Trombetti\,$^{2}$,
C. Burigana\,$^{2,5,6}$, D. Herranz\,$^{7,8}$, M. L\'opez-Caniego\,$^9$, Z.-Y. Cai$^{10}$, L. Bonavera\,$^{11}$ and
J. Gonz\'alez-Nuevo\,$^{11}$}
\def\Address{$^{1}$INAF, Osservatorio Astronomico di Padova, Vicolo dell'Osservatorio 5, I-35122 Padova, Italy \\
$^{2}$INAF, Istituto di Radioastronomia, Via Piero Gobetti 101, I-40129 Bologna, Italy\\
$^{3}$INAF, Italian ALMA Regional Centre, Via Gobetti 101, I-40129, Bologna, Italy \\
$^4$School of Physics and Astronomy, Cardiff University, The Parade, Cardiff CF24 3AA, UK \\
$^5$Dipartimento di Fisica e Scienze della Terra, Universit\`a di Ferrara, Via Giuseppe Saragat 1, I-44122 Ferrara, Italy \\
$^6$INFN, Sezione di Bologna, Via Irnerio 46, I-40127 Bologna, Italy \\
$^7$Instituto de F\'{\i}sica de Cantabria, CSIC-UC, Av. de Los Castros s/n, E-39005 Santander,
    Spain \\
$^8$Departamento de F\'{\i}sica Moderna, Universidad de Cantabria, 39005-Santander,
    Spain \\
$^9$ESAC, Camino Bajo del Castillo s/n, Urb. Villafranca del Castillo 28692, Villanueva de la Canada - Madrid, Spain \\
$^{10}$CAS Key Laboratory for Research in Galaxies and Cosmology, Department of Astronomy,
University of Science and Technology \\of China, Hefei, Anhui 230026, China \\
$^{11}$Departamento de F{\'\i}sica, Universidad de Oviedo, C. Federico Garc{\'\i}a Lorca 18, 33007 Oviedo, Spain}
\def\corrAuthor{Gianfranco De Zotti}

\def\corrEmail{gianfranco.dezotti@inaf.it}

\def\simlt{\mathrel{\rlap{\lower 3pt\hbox{$\sim$}}\raise 2.0pt\hbox{$<$}}}
\def\simgt{\mathrel{\rlap{\lower 3pt\hbox{$\sim$}}\raise 2.0pt\hbox{$>$}}}

\begin{document}
\onecolumn
\firstpage{1}

\title[Running Title]{Extragalactic astrophysics with next-generation CMB experiments}

\author[\firstAuthorLast ]{\Authors} 
\address{} 
\correspondance{} 

\extraAuth{}

\maketitle

\begin{abstract}
\textit{Planck}, SPT and ACT surveys have clearly demonstrated that {Cosmic
Microwave Background (CMB) experiments, while optimised for cosmological
measurements, have made important contributions to the field of extragalactic
astrophysics in the last decade. Future CMB experiments have the potential to
make even greater contributions.} One example is the detection of high-$z$
galaxies with extreme gravitational amplifications. The combination of flux
boosting and of stretching of the images has allowed the investigation of the
structure of galaxies at $z\simeq 3$ with the astounding spatial resolution
of about $60\,$pc. Another example is the detection of proto-clusters of
dusty galaxies at high $z$ when they may not yet possess the hot
intergalactic medium allowing their detection in X-rays or via the
Sunyaev-Zeldovich effect. Next generation CMB experiments, like PICO, CORE,
CMB-Bharat from space and Simons Observatory and CMB-S4 from the ground, will
discover several thousands of strongly lensed galaxies out to $z\sim 6$ or
more and of galaxy proto-clusters caught in the phase when their member
galaxies where forming the bulk of their {stars.  They will also detect tens
of thousands of local dusty galaxies and thousands of radio sources at least
up to $z\simeq 5$. Moreover they will measure the polarized emission of
thousands of radio sources and of dusty galaxies at mm/sub-mm wavelengths.}

\tiny
 \keyFont{ \section{Keywords:} Cosmic Microwave Background, galaxy surveys, radio sources,
 strong lensing, sub-millimeter galaxies, proto-clusters}
\end{abstract}

\section{Introduction}

WMAP and even more \textit{Planck} have already provided an exciting foretaste
of the potential of space-borne Cosmic Microwave Background (CMB) experiments
for extragalactic astrophysics. Next generation experiments with telescope
sizes similar to \textit{Planck\/}'s, like the Cosmic Origins Explorer
\citep[CORE;][]{Delabrouille2018}, the Probe of Inflation and Cosmic Origins
\citep[PICO;][]{Hanany2019} or
CMB\,Bharat\footnote{\url{http://cmb-bharat.in/}} can do much better. This is
because, for state-of-the-art instruments with sensitivity at fundamental
limits (as \textit{Planck} also was) the detection limit is not set by
instrumental noise but by fluctuations of astrophysical foregrounds (confusion
noise), whose amplitude, at the frequencies and angular scales of interest, is
roughly proportional to the beam solid angle \citep[see Fig. 3 of
ref.][]{DeZotti2015}. \textit{Planck} did not work at the diffraction limit
while next generation experiments will. The improvement in angular resolution
will be substantial. For example, at 545 GHz ($550\,\mu$m) the \textit{Planck}
beam has an effective full-width at half maximum $\hbox{FWHM}=4.83'$
\citep{PlanckCollaboration2018HFI} while the diffraction limit for its 1.5\,m
telescope is $1.5'$. Improving the resolution to the diffraction limit
decreases the detection limit at this frequency by about one order of
magnitude.

\begin{figure}
\begin{center}
\includegraphics[width=0.90\textwidth]{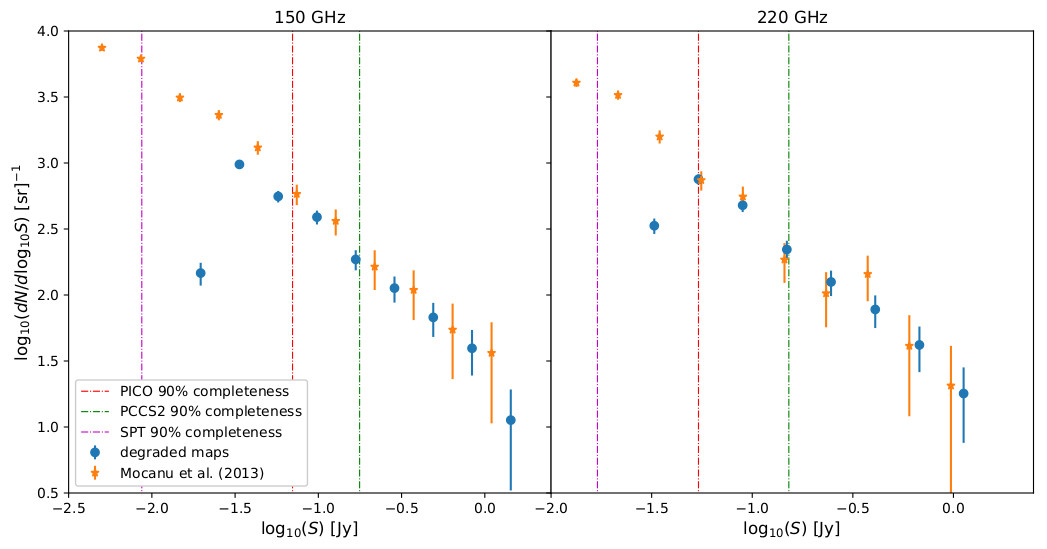}
\end{center}
\caption{{Effect of angular resolution on the confusion limit.}
The filled blue circles show the differential counts of sources on SPT maps degraded to the PICO resolution
at 150\,GHz ($\hbox{FWHM}=6.2'$; left panel) and 220\,GHz ($\hbox{FWHM}=3.6'$; right panel)
compared with the SPT counts of \citet[][orange stars]{Mocanu2013}. The vertical
dot-dashed lines correspond, from left to right, to the 90\% completeness limits
at the full SPT resolution, at the PICO resolution and at the \textit{Planck} resolution.
  }\label{counts_SPT}
\end{figure}

\noindent The \citet{DeZotti2015} results on the impact of angular resolution
on the detection limits, based on simulations exploiting the \textit{Planck}
sky model \citep{Delabrouille2013}, have been improved using real data. To this
end we degraded the publicly available South Pole Telescope (SPT) maps of
$2540\,\hbox{deg}^2$ at 95, 150 and 220\,GHz \citep{Chown2018}, with
resolutions of approximately $1.7'$, $1.2'$, and $1.0'$, respectively, to the
PICO resolution ($\hbox{FWHM}=9.5'$, $6.2'$ and $3.6'$, respectively) and
applied to the degraded maps the Mexican Hat Wavelet\,2 (MHW2) source
detections algorithm \citep{GonzalezNuevo2006, LopezCaniego2006}, also used to
build the Second \textit{Planck} Catalogue of Compact Sources
\citep[PCCS2;][]{PCCS2}. The results at 150 and 220\,GHz are illustrated by
Fig.~\ref{counts_SPT}. The completeness limits for the degraded maps were
obtained by comparison with the SPT counts. We found 90\% completeness down to
95, 70 and 55\,mJy at 95, 150 and 220\,GHz, respectively. For comparison, the
PCCS2 90\% completeness limits\footnote{Effective $\hbox{FWHM}=9.68'$, $7.30'$
and $5.02'$ at 100, 143 and 217\,GHz, respectively; FWHMs of the Gaussians
whose solid angle is equivalent to that of the effective beams.} in the
``extragalactic zone'' are 269, 177 and 142\,mJy, respectively
\citep{PlanckCollaboration2018HFI}. The PICO detection limits at the other
frequencies were obtained by means of analytical extrapolations of results
obtained from simulations done for the CORE project \citep{DeZotti2018}. {These
extrapolations were found} to be consistent with the determinations based on
degraded SPT maps and with the PCCS2 90\% completeness limits in the
``extragalactic zone''. Similar detection limits hold for CORE and CMB Bharat.

\noindent Fluctuations (hence detection limits) at few arcmin resolution are
dominated by compact sources too faint to be detected individually. Such source
confusion can be reliably determined from the power spectra of the Cosmic
Infrared Background (CIB), measured by \textit{Planck}
\citep{PlanckCollaboration2011CIB, PlanckCIB2014} and by \textit{Herschel}
\citep{Viero2013a}. On larger angular scales, however, fluctuations of diffuse
emissions (Galactic or CMB) must also be taken into account.

\begin{figure*}
\begin{center}
\includegraphics[width=0.90\textwidth, trim={0 0 0 0cm}, clip]{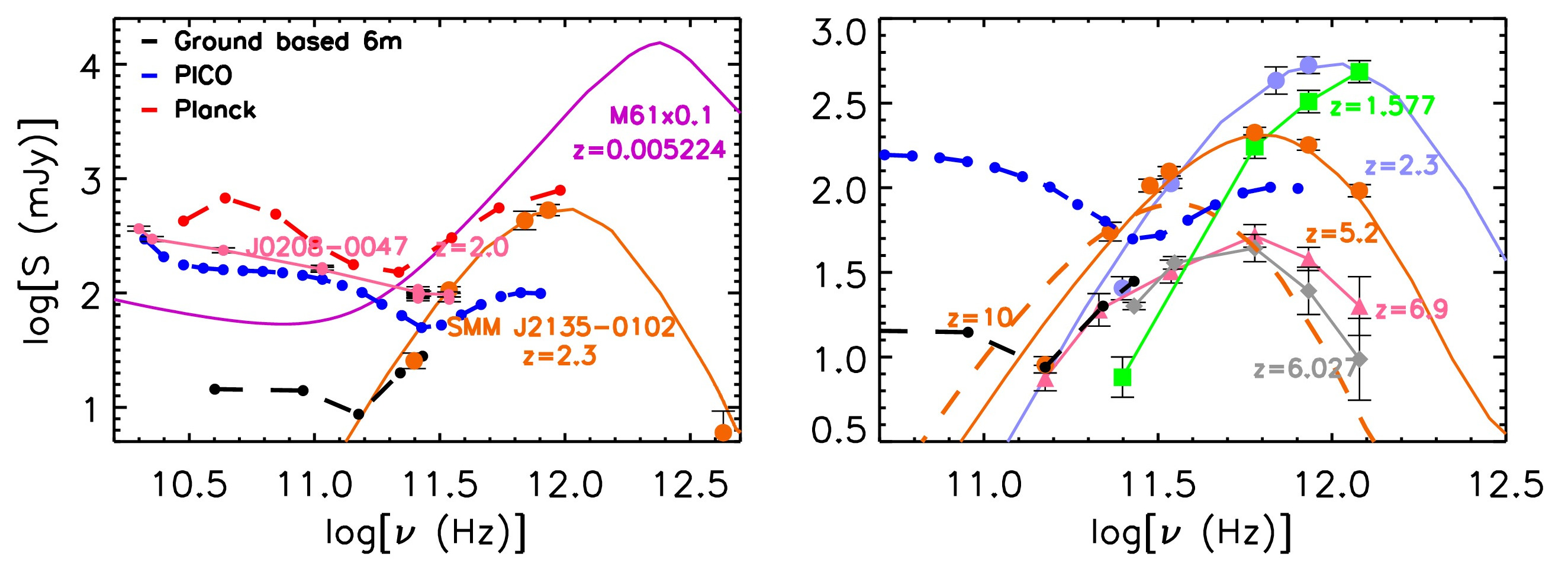}
\end{center}
\caption{Spectral energy distributions of extragalactic sources in the frequency range of CMB
experiments, compared with the estimated detection limits of the PICO project and with the completeness
limits of the SPT surveys extrapolated to 40 and to 270 GHz, to cover the range of the CMB-S4 project.
The PCCS2 90\% completeness limits in the ``extragalactic zone'' are also shown for comparison (in the left panel only).
At wavelengths shorter than a few mm the dominant extragalactic population are dusty star-forming
galaxies. At longer wavelengths radio sources, mostly blazars,  take over. {The spectral energy
distributions (SEDs) of blazars are characterized by a relatively flat continuum spectrum exemplified by
that of J0208-0047 at $z\simeq 2.0$ (flat solid orange line).}
Dusty star-forming galaxies are a mixture of local and high-$z$ strongly gravitationally lensed objects.
{Local galaxies are, by far, the brightest extragalactic sources at sub-mm wavelengths. Their SEDs
 are exemplified by that of the dusty star-forming galaxy M\,61, scaled down by a factor of 10.}
Some examples of SEDs of strongly lensed galaxies are shown: SDP\,9 \citep[$z=1.577$;][]{Negrello2010},
SMMJ2135-0102 \citep[$z=2.3259$;][]{Swinbank2010}, HLS\,J091828.6+514223  \citep[$z=5.2$;][]{Combes2012},
HATLAS\,J090045.4+004125 \citep[$z=6.027$;][]{Zavala2018} and SPT-S\,J031132-5823.4 \citep[$z=6.9$;][]{Strandet2017}.
The dashed orange SED represents that of HLS\,J091828.6+514223 scaled to $z=10$ to show that galaxies like
it are visible up to very high redshifts both by ground-based and by space-borne next generation CMB experiments.
} \label{fig:SEDs}

\end{figure*}

\noindent The potential of ground based CMB experiments has been demonstrated
by the SPT \citep{Mocanu2013} and by the Atacama Cosmology Telescope
\citep[ACT;][]{Marsden2014} surveys. With telescopes of the 6--10\,m class,
these experiments reach arcmin angular resolution at mm wavelengths, with
sensitivity at fundamental limits. Next generation experiments, like CMB-S4
\citep{Abazajian2016} and the Simons Observatory \citep{Ade2019} will extend
the coverage to at least 40\% of the sky.

\noindent Space-borne experiments like PICO will cover a very broad frequency
range, from $\simeq 20$ to $\simeq 800\,$GHz, while observations from the
ground are only possible, in atmospheric windows, up to $\simeq 300\,$GHz. The
extragalactic sky in the frequency range of CMB experiments is dominated by two
point source populations: blazars (Flat-Spectrum Radio Quasars and BL
Lacertae-type sources) and dusty star-forming galaxies. Blazars {are} powerful
radio-loud active galactic nuclei (AGNs) whose relativistic jets are closely
aligned with the line of sight. {They} overbear the number counts at
wavelengths longer than about 1\,mm {while} dusty galaxies prevail at shorter
wavelengths. This is illustrated by Fig.~\ref{fig:SEDs}, where the spectral
energy distributions (SEDs) of a $z\simeq 2$ blazar and of two dusty galaxies
are compared with the estimated detection limits of the PICO project
\citep{Hanany2019} and with the completeness limits of the SPT surveys
\citep{Mocanu2013}, extrapolated on one side to 40\,GHz and on the other side
to 270\,GHz to cover the frequency range of CMB-S4. The 90\% completeness
limits of the PCCS2 in the ``extragalactic zone'' are also shown for
comparison.

\noindent In this paper we will discuss the expected outcome of planned next
generation CMB experiments with regard to strongly gravitationally lensed
high-$z$ dusty galaxies (Sect.~\ref{sect:lensed}), galaxy proto-clusters
(Sect.~\ref{sect:protocl}) and radio sources (Sect.~\ref{sect:radio}). In
Sect.~\ref{sect:counts_pol} we deal with counts of the various classes of
extragalactic sources in polarization. The main conclusions are summarized in
Sect.~\ref{sect:conclusions}

\noindent We adopt a flat $\Lambda$CDM cosmology with the latest values of the
parameters derived from Planck CMB power spectra: $H_0 =
67.4\,\hbox{km}\,\hbox{s}^{-1}\, \hbox{Mpc}^{-1}$ and $\Omega_m = 0.315$
\citep{PlanckParameters2018}.

\begin{figure*}[h!]
\begin{center}
\includegraphics[width=0.90\textwidth, trim={0 0 0 0cm}, clip]{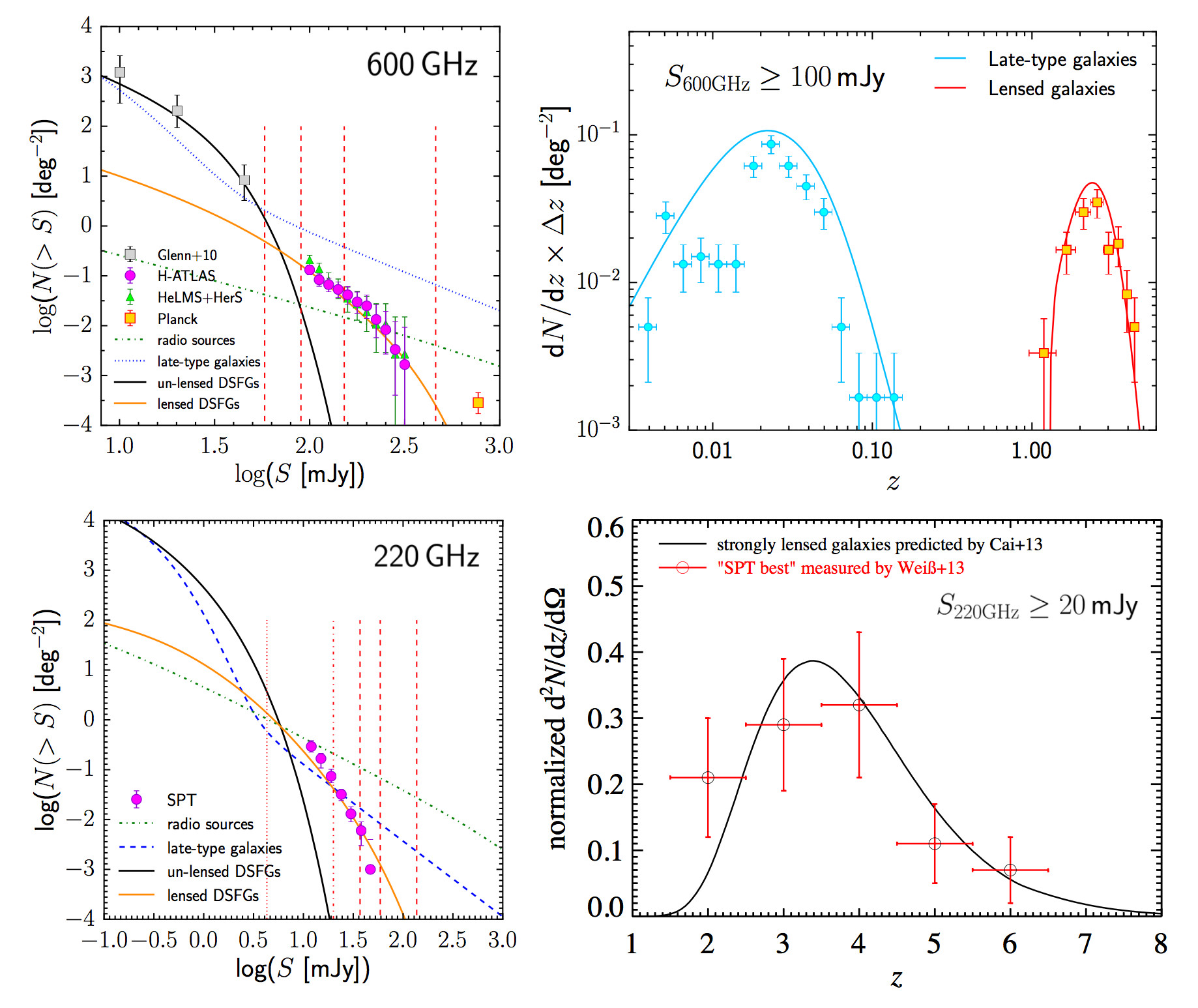}
\end{center}
\vskip-0.7cm
\caption{\textbf{Upper left panel.} Integral counts of strongly lensed galaxies from
\textit{Herschel} surveys at $500\,\mu$m \citep[600 GHz; green and magenta data
points from ref.][]{Negrello2017lensed}, compared with the predictions of the \citet{Cai2013} model
(solid orange line). The yellow square on the bottom-right
corner is our own estimate of the counts of strongly lensed galaxies detected
by \textit{Planck}. The counts of unlensed proto-spheroidal galaxies
(data points from ref. \citep{Glenn2010}, black squares; model by ref. \citep{Cai2013}, solid black line),
of radio sources \citep[dotted green line, model by ref.][]{Tucci2011} and of normal and starburst late-type
galaxies \citep[dotted blue line, model by ref.;][]{Cai2013} are also shown for comparison. The
vertical dashed orange lines show, from right to left, the 90\% completeness
limit of the PCCS2 and the $5\,\sigma$ detection limits for space-borne CMB
experiments with 1, 1.5 and 2\,m telescopes working at the diffraction limit.
This panel shows a modified version of Fig.~8 of \citet{Negrello2017lensed}.
\textbf{Upper right panel.} {Redshift distribution of galaxies brighter than 100\,mJy
at $500\,\mu$m, derived from the full H-ATLAS catalogue (data points with Poisson errors)
compared with the predictions of the \citet{Cai2013} model (solid lines).} There is a clear
bimodality. On one side we have nearby late-type galaxies, almost all at $z\le
0.06$, and hence easily recognizable in optical/near-infrared catalogues. On
the other side we have dust enshrouded, hence optically very faint,
gravitationally lensed galaxies at $z\ge 1$ and up to $z>4$. \textbf{Lower left panel.}
Integral counts of strongly lensed galaxies from the SPT survey at
$220\,$GHz \citep[1.4\,mm; magenta data points;][]{Vieira2010} compared with the
prediction of the \citet[][orange solid line]{Cai2013} model. Counts of radio sources
and of late-type galaxies are from the same models as in the upper panel. The vertical
lines show, from right to left, the $5\,\sigma$ detection limits for diffraction limited
1, 1.5 and 2\,m telescopes, the completeness limit of the SPT survey \citep{Mocanu2013} and
the $5\,\sigma$ confusion limit for a 6\,m telescope.
\textbf{Lower right panel.} Estimated redshift distribution of strongly lensed galaxies with
$S_{220\,\rm GHz}>20\,$mJy \citep{Weiss2013} compared with the prediction of the \citet{Cai2013} model.}
\label{fig:lensed_Herschel_SPT}
\end{figure*}

\section{Dusty galaxies}\label{sect:lensed}

\subsection{Strongly-lensed galaxies}

\noindent Next generation CMB experiments will conduct a census of the
brightest sub-mm galaxies in the Universe. \textit{Planck}, SPT and ACT have
already offered an exciting foretaste of that. Follow-up CO spectroscopy and
multi-frequency photometry of 11 ``\textit{Planck}'s dusty GEMS''
\citep[Gravitationally Enhanced subMillimetre Sources;][]{Canameras2015} have
shown that they are at $z=2.2$--3.6. {Their apparent (uncorrected for
gravitational amplification) far-IR  luminosities are} up to $3\times
10^{14}\,\hbox{L}_\odot$, making them among the brightest sources in the
Universe.

\noindent \textit{Herschel} extragalactic surveys yielded a surface density of
$\simeq 0.16\,\hbox{deg}^{-2}$ for $S_{500\mu\rm m} \ge 100\,$mJy
\citep{Negrello2017lensed}. A space-borne 1.5\,m telescope will reach a
slightly fainter flux density limit over the full sky (excluding the region
around the Galactic plane), thus achieving the detection of several thousands
of strongly lensed galaxies (upper left-hand panel of
Fig.~\ref{fig:lensed_Herschel_SPT}). The number of detections decreases rapidly
for smaller telescope sizes; a telescope substantially smaller than 1\,m cannot
do any better than \textit{Planck}. A 2\,m telescope would increase the number
of strongly lensed detections by a factor $\simeq 2.5$. {However} their
selection would require additional information to distinguish them from
unlensed high-$z$ galaxies dominating the counts at the corresponding flux
density limit. Efficient methods to separate lensed and unlensed galaxies have
been presented by refs.~\citep{GonzalezNuevo2012, GonzalezNuevo2019}.

\noindent  {Ground-based instruments typically have higher resolution at the
longer wavelengths, and thus select a higher redshift population.} A
straightforward extrapolation of results from analyses in the first
\citep{Vieira2010} and second \citep{Mocanu2013} SPT catalogue papers imply
that the SPT has detected hundreds of strongly lensed galaxies over the full
survey area (lower left-hand panel of Fig.~\ref{fig:lensed_Herschel_SPT}).
Their redshift distribution \citep{Weiss2013, Vieira2013} is broader than found
for \textit{Herschel} surveys \citep{Negrello2017lensed} and the mean redshift,
$\bar{z}=3.5$ \citep{Weiss2013} is higher (lower right-hand panel of
Fig.~\ref{fig:lensed_Herschel_SPT}). The maximum spectroscopically measured
redshift, $z=6.9$ \citep{Strandet2017}, shows that this survey has reached the
epoch of reionization. Additional samples of strongly lensed galaxies, over a
similar redshift range have been provided by the ACT surveys
\citep{Marsden2014, Su2017}.

\noindent Next-generation ground-based experiments, extending the sky coverage
up to 40\% of the sky or more, will detect thousands of strongly lensed
galaxies if their performances will be similar to those of the SPT. If these
experiments will be able to reach the confusion limit of a 6\,m telescope
operating at the diffraction limit (4-5\,mJy), the number of strongly-lensed
detections will increase to tens of thousands.

\noindent The right-hand panel of Fig.~\ref{fig:SEDs} shows that both
ground-based and space-borne experiments will detect strongly lensed galaxies
up to high redshifts. For example objects like the strongly lensed galaxy
HLS\,J091828.6$+$514223 at $z=5.2$  would be detectable up to $z$ of at least
10 (dashed SED in Fig.~\ref{fig:SEDs}). Ground-based and space-borne
experiments cover complementary redshift ranges. The former are more efficient
at $z\simgt 2$, the latter at lower $z$.

\noindent These surveys will also probe a large fraction of the entire Hubble
volume for the most intense hyper-luminous starbursts, testing whether there
are physical limits to the star-formation rates (SFRs) of galaxies \citep[e.g.,
ref.][]{Crocker2018}.

\begin{figure}[h!]
\begin{center}
\includegraphics[width=10cm]{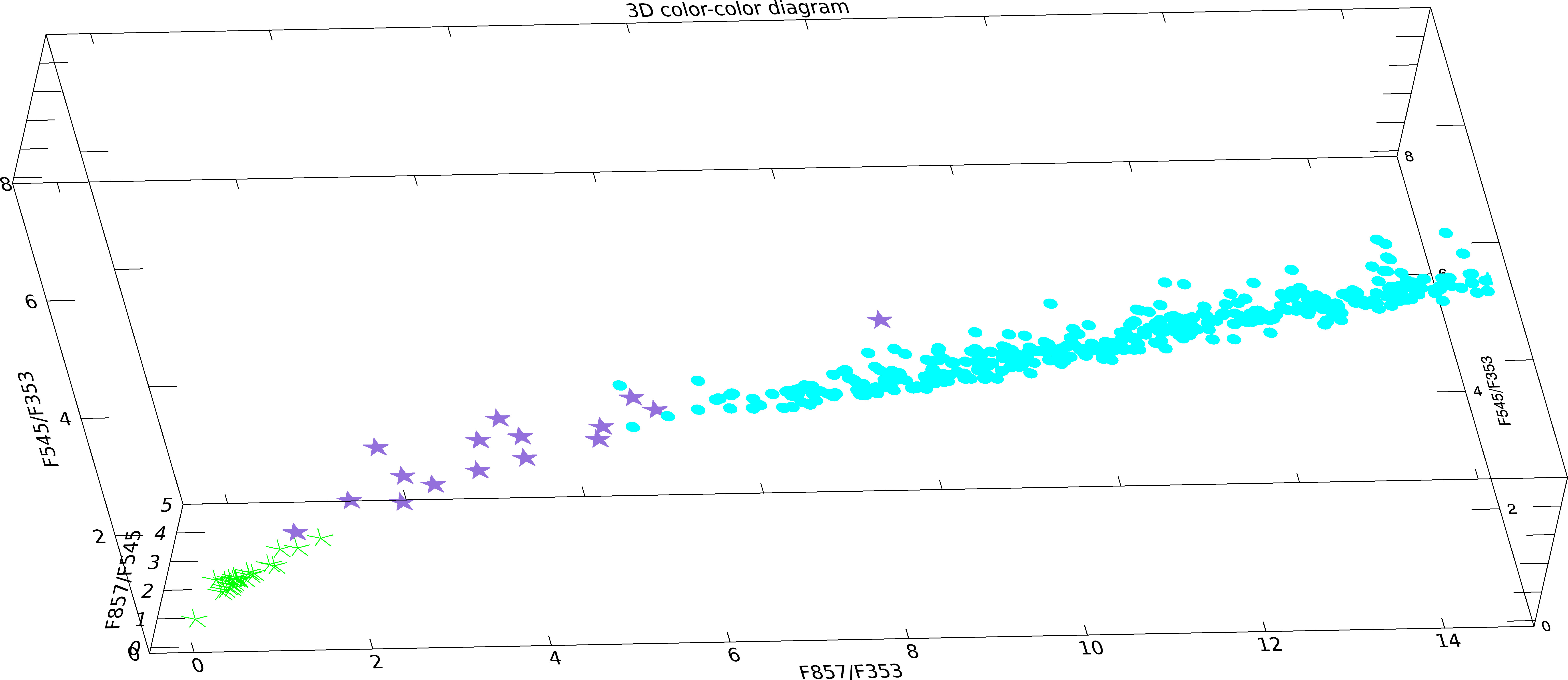}
\end{center}
\caption{3D colour-colour plot showing the colours of strongly-lensed galaxies detected
by \textit{Planck} (purple stars) compared to those of local galaxies (light blue symbols)
and of radio sources (green symbols). The distribution of local galaxies extends to the blue
far beyond the chosen boundaries of the figure. Strongly lensed galaxies populate a region
intermediate between those of local galaxies and radio sources and well distinct from both.}\label{fig:3Dcolours}
\end{figure}

\begin{figure}[h!]
\begin{center}
\includegraphics[width=10cm]{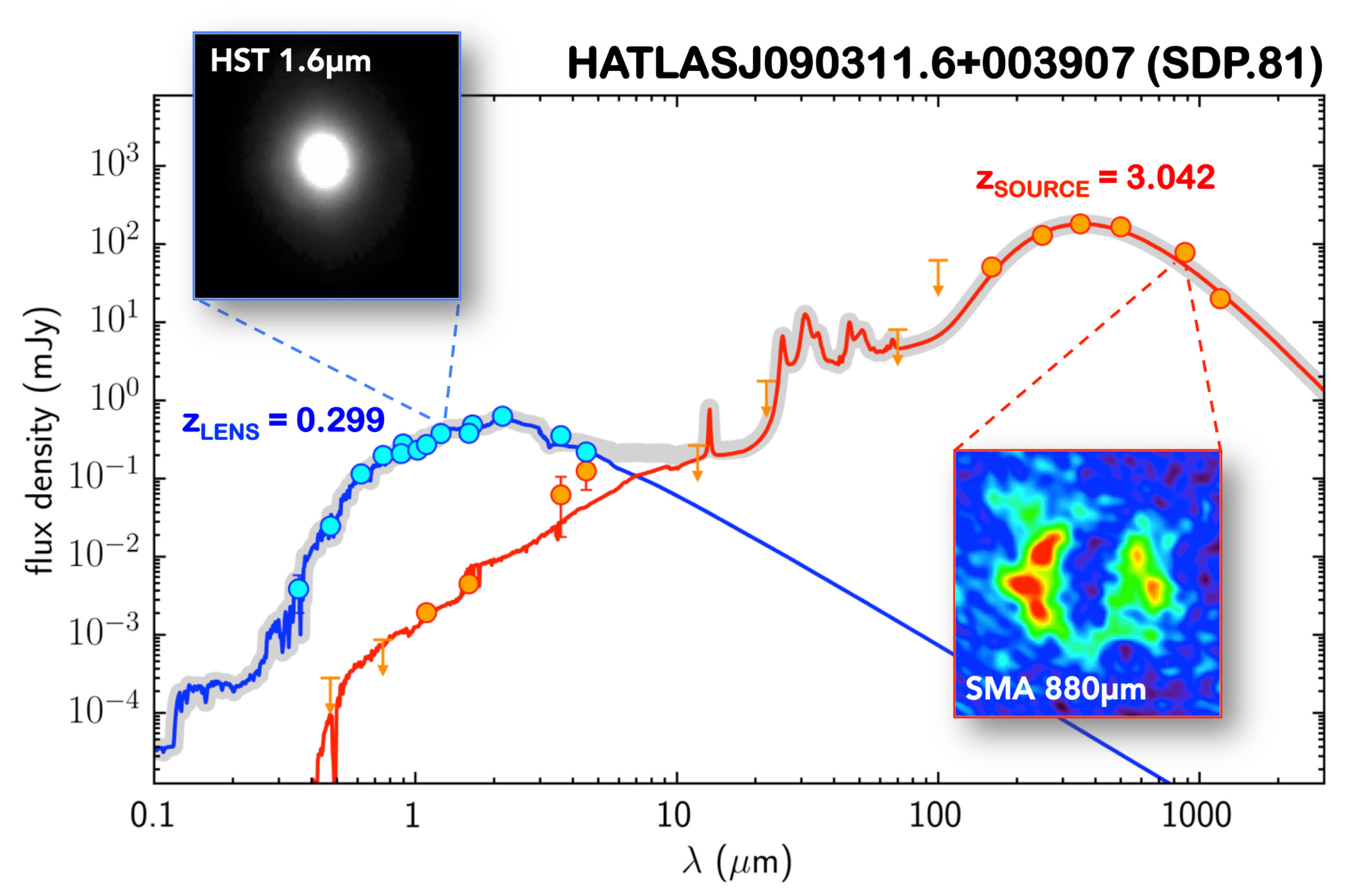}
\end{center}
\caption{SEDs of the strongly lensed galaxy SDP\,81 at $3.042$ detected by the \textit{Herschel} Astrophysical
Terahertz Large Area Survey \citep[H-ATLAS;][]{Eales2010} and imaged
with the Sub-Millimeter Array (SMA) at $880\,\mu$m \citep{Negrello2010} and of the foreground lens
at $z=0.299$ imaged with the Wide Field Camera\,3 (WFC3) on the Hubble Space Telescope
\citep[HST;][]{Negrello2014}. As is generally the case, the lens is a spheroidal galaxy in passive
evolution, hence very faint at sub-mm wavelengths; indeed it is essentially invisible in the SMA image.
The lensed source is dust-enshrouded, hence bright in the sub-mm but very faint in the optical and
essentially invisible in the WFC3/HST image.
 }\label{fig:SDP81}
\end{figure}

\subsection{Comparison with the selection of strongly lensed galaxies in other wavebands}

\noindent \textit{Herschel} surveys have demonstrated that at $500\,\mu$m
(600\,GHz) strongly lensed galaxies with $S_{500\mu\rm m}\ge 100\,$mJy, close
to the detection limit of next generation space-borne CMB experiments with
telescopes of the $\simeq 1.5\,$m class operating at the diffraction limit,
amount to $\simeq 25\%$ of the total counts \citep[][cf. upper left panel of
Fig.~\ref{fig:lensed_Herschel_SPT}]{Negrello2017lensed, Nayyeri2016}. A similar
fraction was found at the completeness limit ($\simeq 20\,$mJy) of the SPT
survey at 220 GHz \citep[][]{Mocanu2013}. Such high fractions {are a direct
consequence of the extreme steepness of (sub-)mm counts of extragalactic
sources due to the unique combination of steep cosmological evolution with a
strongly negative K-correction\footnote{{The steep increase with increasing
frequency of the dust emission spectrum at (sub-)mm wavelengths ($L_\nu \propto
\nu^\alpha$ with $\alpha \simeq 3.5$--4) translates into a fast increase of the
observed flux density with increasing redshift (negative K-correction). Such
increase may compensate and even exceed the decrease due to increasing distance
\citep{Franceschini1991, BlainLongair1993}} }. They} are an exclusive property
of (sub-)mm surveys: searches in other wavebands have yielded fractions of
$\simeq 0.1\%$ \citep{York2005, Jackson2008, Treu2010}.

\noindent Not only the strongly-lensed fractions are high, but also the
selection is easy done, with close to 100\% efficiency \citep{Negrello2007,
Negrello2010}, while it remains a significant challenge in other wavebands
\citep{Jacobs2019}. The other extragalactic sources brighter than the detection
limits of next generation CMB experiments are either $z\simlt 0.1$ galaxies or
radio sources (mainly blazars), that can be identified using existing all-sky
optical or radio catalogs.

\noindent In the case of space-borne experiments, the selection can be done
directly on survey data. As illustrated by the upper right-hand panel of
Fig.~\ref{fig:lensed_Herschel_SPT}, the redshift distribution of dusty galaxies
detected by these experiments is highly bimodal. On one side we have nearby
late-type galaxies, almost all at $z\le 0.06$;  on the other side we have dust
enshrouded, hence optically very faint, gravitationally lensed galaxies at
$z\ge 1$ and up to $z>4$. This strong difference in redshift translates in a
clear difference in sub-mm colours, high-$z$ galaxies being much redder than
the local ones; radio sources have still redder colours (see
Fig.~\ref{fig:3Dcolours}). This is specific to searches with these experiments.
Selections in other wavebands need spectroscopy or other ancillary data. This
is true also for ground-based surveys which provide photometry in the
Rayleigh-Jeans region where colours are largely redshift-independent. {On the
other hand, the discrimination of strongly lensed galaxies will be greatly
eased by the wealth of multi-wavelength data over the entire sky that will be
available in the coming decade.}

\noindent In addition to the much more efficient selection, other critical
advantages of CMB experiments over other facilities that will generate large
gravitational lens catalogues \citep[e.g., \textit{Euclid}, Gaia,
SKA;][]{Serjeant2017} are:
\begin{itemize}
\item compared to optical/near-IR surveys, they detect earlier (higher
    redshift) phases of galaxy evolution, characterized by intense,
    dust-enshrouded star-formation activity;
\item the photometry of detected galaxies will be only very weakly
    contaminated by the foreground lens, even in the case of close alignment
    along the line-of-sight (see Fig.~\ref{fig:SDP81}); while the lensed
    galaxies are heavily dust enshrouded, hence bright at sub-mm wavelengths
    but faint in the optical, the foreground lenses are mostly massive
    early-type galaxies in passive evolution, hence optically bright but
    almost invisible in the sub-mm;
\item the all-sky coverage maximizes the detections of the rare brightest
    sources, with the most extreme magnifications, optimally suited for
    follow-up, as demonstrated by \textit{Planck}: the magnification factors
    of ``\textit{Planck}'s dusty GEMS'' are estimated to be of up to 50
    \citep{Canameras2015};
\item the mm/sub-mm selection, with its strongly negative K-correction,
    allows us to extend the detection of sources and lenses to much higher
    redshifts than in any other waveband.
\end{itemize}

\subsection{Astrophysics and cosmography with strong lensing}

\noindent The extreme magnifications of dusty galaxies detected by CMB
experiments makes them trivially easy targets for ALMA, NOEMA, SMA etc., and
the foreground lenses will almost certainly be detectable by large optical
telescopes and in, e.g., \textit{Euclid} imaging. Follow-up observations will
allow us to address major astrophysical issues \citep[e.g.,][]{Treu2010}.

\noindent CMB experiments with arcmin resolution both from the ground and from
space will drive a real breakthrough in the study of early evolutionary phases
of galaxies,  paving the way to answer major, still open problems like: which
are the  physical mechanisms shaping the galaxy properties
\citep{SilkMamon2012, SomervilleDave2015}: in situ processes? interactions?
mergers? cold flows? How feedback processes work? To settle these issues we
need direct information on the structure and the dynamics of high-$z$ galaxies.
But these are compact, with typical sizes of 1--2\,kpc
\citep[e.g.,][]{Fujimoto2018, Enia2018}, corresponding to angular sizes of
0.1--0.2\,arcsec at $z=2$--3. Thus they are hardly resolved even by ALMA and by
the HST. If they are resolved, high enough signal-to-noise ratios per
resolution element are achieved only for the brightest galaxies, not
representative of the general population.

\noindent Strong gravitational lensing provides a solution. CMB surveys will
detect  the brightest (sub-)mm strongly lensed galaxies in the sky, with
extreme magnifications, up to several tens \citep{Canameras2015}. Since lensing
conserves surface brightness, the effective angular size is stretched by an
average factor $\mu^{1/2}$, substantially increasing the resolving power. A
spectacular example are ALMA 0.1 arcsec resolution observations of PLCK\,G244.8
+54.9 at $z\simeq 3.0$ with $\mu \simeq 30$ \citep{Canameras2017ALMA} which
reached the astounding spatial resolution of $\simeq 60\,$pc, substantially
smaller than the size of Galactic giant molecular clouds.

\noindent \citet{Canameras2017ALMA} also obtained CO spectroscopy with an
uncertainty of 40—50 km/s. This spectral resolution makes possible a direct
investigation of massive outflows driven by AGN feedback at high $z$, with
predicted velocities of $\sim 1000\,\hbox{km}\,\hbox{s}^{-1}$
\citep{KingPounds2015}. AGN driven outflows are a key ingredient of current
galaxy evolution models since they provide the most plausible explanation for
the deviation of the galaxy stellar mass function from the halo mass function
at large masses, i.e. for the low star-formation efficiency in massive halos:
only less than 10\% of baryons initially present in such halos are used to form
stars. However, until very recently, with few exceptions \citep{Cicone2015},
high-$z$ outflows were detected only in ionized gas \citep{CresciMaiolino2018}.
Thus information on the effect of feedback on the direct fuel of star
formation, molecular gas, has been largely missing during the epoch of the most
active cosmic star formation.

\noindent {Due to the weakness of spectral signatures of molecular outflows,
observational progress has been difficult. Gravitational lensing allows us to
overcome these difficulties.} \citet{Spilker2018} were able to detect, by means
of ALMA spectroscopy,  a fast (800\,km/s) molecular outflow in a strongly
lensed galaxy at $z=5.3$, discovered by the SPT survey. The outflow carries
mass at a rate close to the SFR, thus removing a large fraction of the gas
available for star-formation. \citet{Canameras2018outflow} detected a molecular
wind signature in the strongly lensed galaxy PLCK\,G165.7+49.0, discovered by
\textit{Planck},  at $z=2.236$, with magnification factors between 20 and 50 in
most of the source. Strongly lensed galaxies detected by CMB experiments will
be obvious targets for the next generation Very Large Array (ngVLA) that will
provide accurate estimates of the molecular outflow masses and mass loss rates
up to $z\simeq 4$ \citep{SpilkerNyland2018}. The proposed NASA flagship Origins
Space Telescope \citep[OST;][]{Leisawitz2018, Bonato2019} would detect
molecular, neutral, and warm ionized phases of outflows up to higher redshifts,
although with a spatial resolution lower than ngVLA. AGN-driven outflows also
produce a bubble of hot gas, potentially detectable via its Sunyaev-Zeldovich
effect \citep[SZE;][]{NatarajanSigurdsson1999, Platania2002}. The first
detection of the SZE from a hyperluminous quasar at $z=1.71$, obtained by means
of ALMA observations, was reported by \citet{Lacy2019} who, from these
measurements, derived constraints on the energetics of the wind.

\noindent The high redshifts of magnified galaxies imply high redshifts of
foreground lenses. Optical follow-up allows us to investigate the total
(visible and dark) mass of the lensing galaxies, their density profiles, dark
matter sub-structures at higher redshifts than in the case of optical
selection. For example, the spectroscopic redshift of the foreground deflector
of the strongly lensed galaxy PLCK\,G244.8+54.9 ($z=3.005$), measured by
\citet{Canameras2017lens}, {was found to be exceptionally high, $z = 1.525$.
This illustrates the power of the (sub-)mm selection of strongly lensed
galaxies to push the study of dark matter and baryon assembly up to high
redshifts.}

\noindent The samples of thousands of strongly lensed galaxies provided by CMB
experiments will also be a powerful tool to measure cosmological parameters
\citep{Treu2010, Eales2015}. The lens equation contain ratios of angular
diameter distances. Hence observables of lensed systems (redshifts of the
source and of the lens, Einstein radius, velocity dispersion of the lens) can
be used to estimate $\Omega_m$, $\Omega_\Lambda$ and the parameters of the dark
energy equation of state. These determinations will not be as precise as those
obtained from CMB data, but will provide valuable independent tests of the
cosmological model, will allow to look for unrecognized selection effects and
to break degeneracies in the interpretation of CMB data. Also, the strong
lensing optical depth depends on the abundance and redshift distributions of
potential wells acting as deflectors, hence on density parameters, on the dark
energy equation of state and its evolution and on the amplitude, $\sigma_8$, of
the primordial perturbation spectrum. Distortions of Einstein rings or giant
arcs could offer a direct method to detect sub-halos, i.e. to constrain warm
dark matter \citep{Li2016}. Lenses at high redshift are particularly valuable
to determine the statistics of perturbations along the line of sight
{\citep{Treu2010}}.

\subsection{Local dusty galaxies}

\noindent \textit{Herschel} surveys have shown that the surface density of
local dusty galaxies brighter than $S_{500\mu\rm m}=100\,$mJy is $\simeq
1\,\hbox{deg}^{-2}$ and that their redshift distribution peaks at
$z=0.02$--0.03 \citep[upper right panel of
Fig.~\ref{fig:lensed_Herschel_SPT};][]{Negrello2017lensed}. A survey with a
space borne $\simeq 1.5\,$m telescope will detect tens of thousands of them. By
the time when next generation CMB experiments will fly several wide-angle
redshift and photometric surveys will be available, providing distance
information for the majority, if not all of them \citep{DeZotti2018}. Combining
these data with available or forthcoming data in different wavebands (radio,
IRAS, AKARI, WISE, Euclid, GALEX, ROSAT, eROSITA ...) it will be possible to
determine, for each galaxy type and as a function of stellar mass, the
distribution of dust temperatures and masses, the SFR function, the
relationship between star formation and nuclear activity, the contributions of
newborn and evolved stars to dust heating, and more. The sample of local
galaxies will also be large enough for clustering studies, i.e. to relate the
properties of galaxies to the underlying dark matter field and to the
properties of their dark matter haloes, as well as to investigate the link
between galaxies of different types and their environments.

\noindent The surface density of local galaxies at the SPT completeness limit
at 220\,GHz (20\,mJy) is of $\simeq 0.05\,\hbox{deg}^{-2}$ (lower {left} panel
of Fig.~\ref{fig:lensed_Herschel_SPT}), i.e. a factor of $\simeq 20$ lower than
achieved by a space-borne 1.5\,m telescope at 600\,GHz. However, as shown by
the SED of M\,61 scaled down by a factor of 10 (Fig.~\ref{fig:SEDs}), ground
based surveys will detect, for the nearest low-$z$ galaxies, also the radio
emission powered by star formation. At mm wavelength such radio emission is
expected to be dominated by free-free emission \citep{Galvin2018} which is an
excellent measure of the instantaneous SFR since it scales with the ionizing
luminosity and is unaffected by extinction. Its measurement will allow us to
recalibrate the relation between dust emission (which may be contributed in
part by dust heating from old stars) and SFR. According to the calculations by
\citet{Mancuso2015}, the surface density of star-forming galaxies detected via
their radio emission at 95\,GHz by SPT-like experiments, i.e. brighter than
$S_{95\rm GHz}\simeq 10\,$mJy, is of $\simeq 12\,\hbox{sr}^{-1}$. {We caution,
however, that the measurement of the free-free emission requires the separation
of the non-thermal (synchrotron) and dust contributions. This is very difficult
to do accurately, especially because of uncertainties on the synchrotron
spectrum \citep{Aravena2013}.}

\noindent Another open issue is the connection between nuclear radio activity
and star formation. \textit{Planck} has detected {evidence} of cold dust
associated to a handful of nearby radio sources, based on their rising spectra
at mm wavelengths \citep{PlanckCollaborationXLV2016}. The much deeper surveys
carried out by next-generation experiments can push the investigation to much
more distant objects.

\begin{figure}[h!]
\begin{center}
\includegraphics[width=8cm]{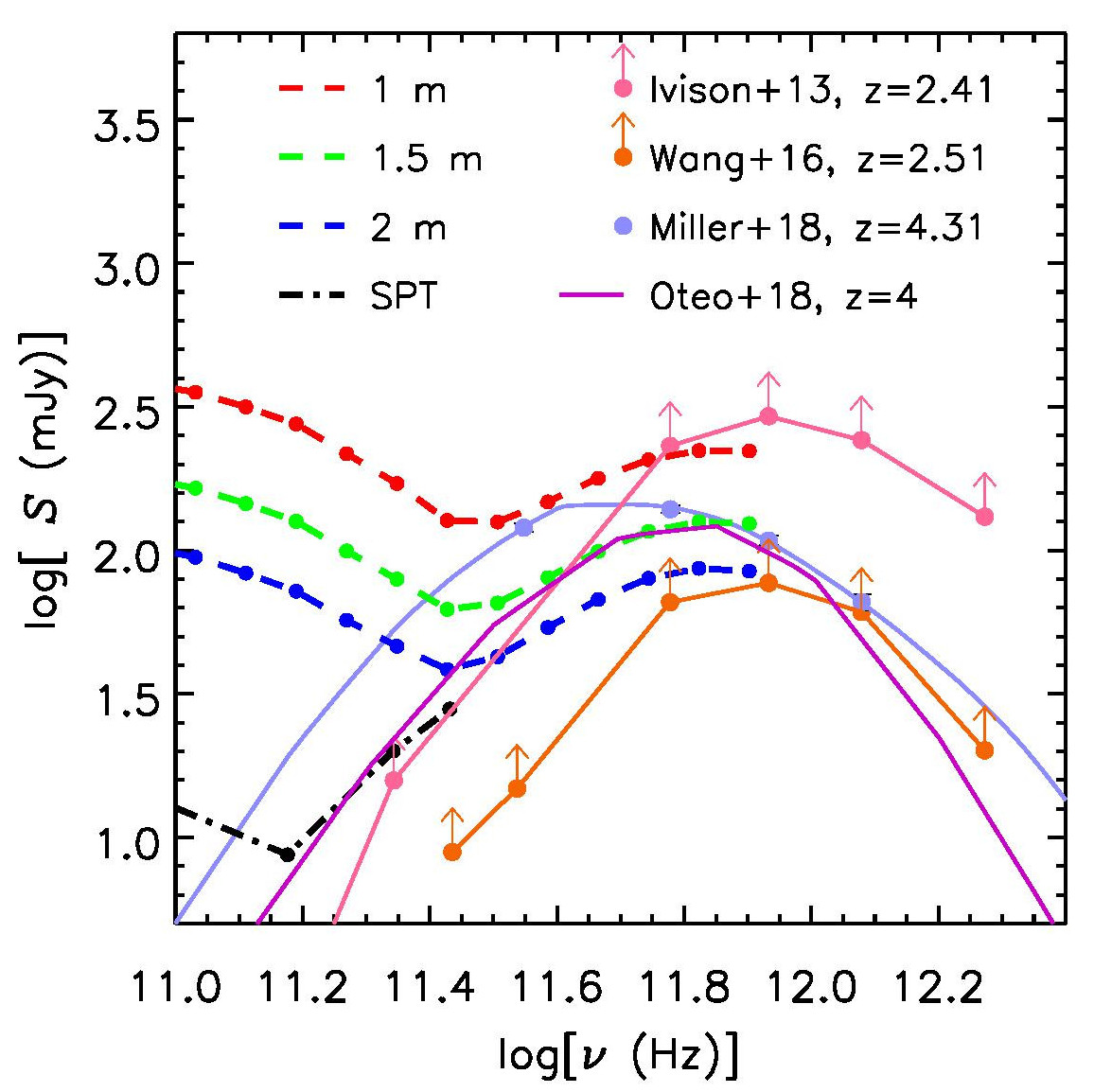}
\end{center}
\caption{SEDs of spectroscopically confirmed sub-mm bright high-$z$ proto-clusters
detected by \citet{Ivison2013}, \citet{Wang2016}, \citet{Miller2018} and \citet{Oteo2018},
compared with (from top to bottom) the detection limits of diffraction limited
space-borne telescopes of 1, 1.5 and 2\,m size, and with the SPT completeness limit.
The flux densities by \citet{Ivison2013} and \citet{Wang2016} are shown as lower
limits because they include the contributions of member galaxies across a $\sim 100\,$kpc region,
substantially smaller than the expected proto-cluster size ($\sim 500\,$kpc).}\label{fig:SEDprotocl}
\end{figure}

\begin{figure}
\begin{center}
\includegraphics[width=\textwidth, trim={0 0 0 0cm}, clip]{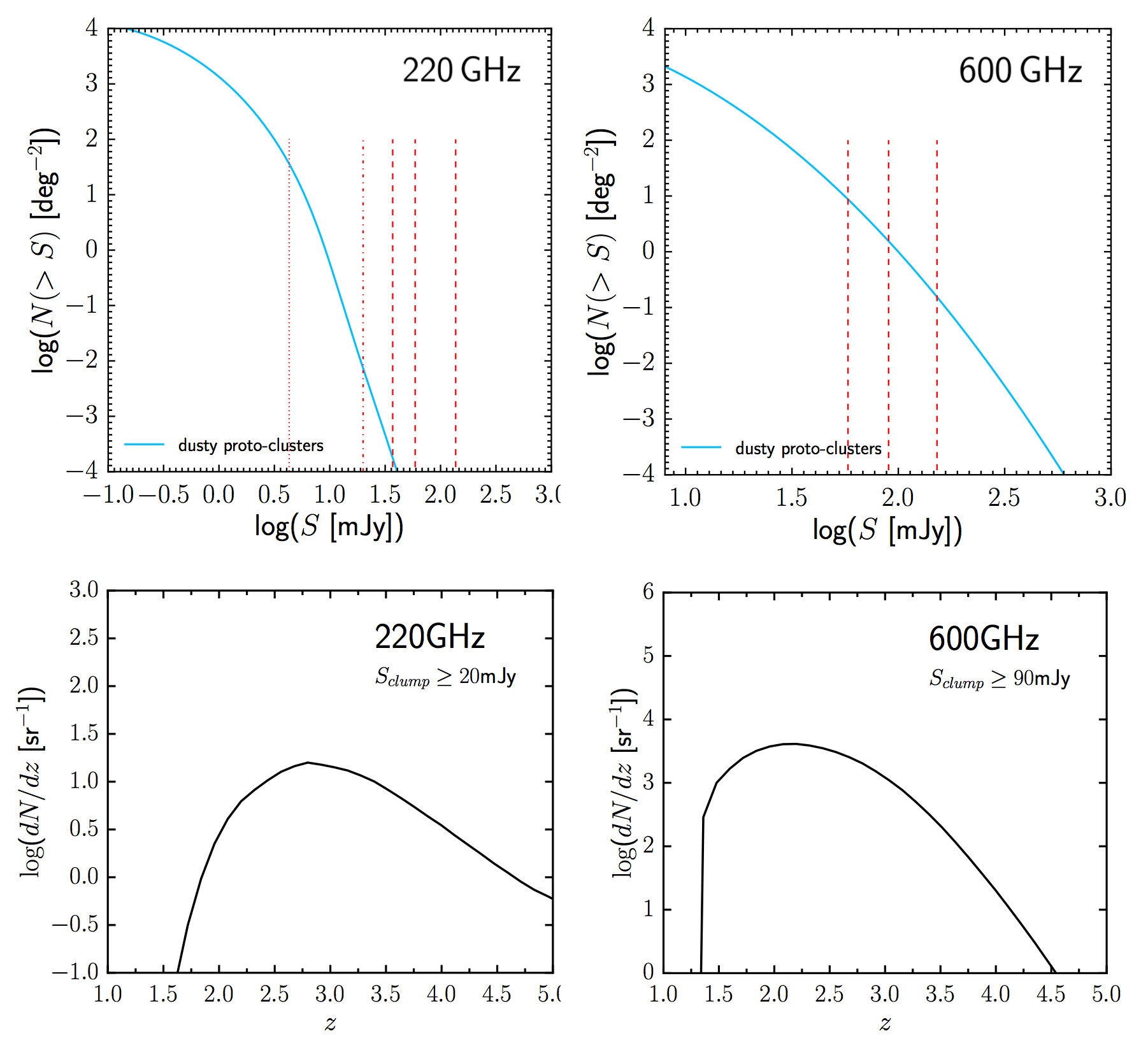}
\end{center}
\caption{Integral counts and redshift distributions of proto-clusters at 220\,GHz and 600\,GHz
(1.4\,mm and $500\,\mu$m, respectively) predicted
by \citet{Negrello2017protocl}. The vertical lines in the upper left panel show, from left to right,
the $5\,\sigma$ confusion limit for a 6\,m telescope, the SPT completeness limit and the $5\,\sigma$
detection limits for 2, 1.5 and 1\,m telescopes operating at the diffraction limit. Only the three latter limits,
at $600\,$GHz, are shown in the upper right-hand panel. The lower panels show the redshift distributions at the SPT
completeness limit at 220\,GHz (20\,mJy, left) and at the $5\,\sigma$ detection limit (90\,mJy) for a 1.5\,m telescope
at 600\,GHz (right). }\label{fig:protocl_counts}
\end{figure}


\section{Early phases of cluster evolution}\label{sect:protocl}

Understanding the full evolutionary history of  present-day  galaxy clusters is
of fundamental importance for the observational validation of the formation
history of the most massive dark matter halos, a crucial test of models for
structure formation, and for investigating the impact of environment on the
formation and evolution of galaxies. Because of their deep potential wells,
clusters may preserve fingerprints of the physical processes responsible for
triggering and suppression of star formation and black hole activity. Also,
clusters of galaxies have historically been powerful probes of cosmological
parameters \citep[e.g., ref.][]{Bocquet2018}.

\noindent This has driven extensive searches for high redshift galaxy
proto-clusters in the past two decades \citep[for a review see
ref.][]{Overzier2016}. Classical systematic cluster searches were carried out
via X-ray \citep[e.g., refs.][]{Mehrtens2012, Willis2013}, Sunyaev-Zeldovich
\citep[SZ;][]{SunyaevZeldovich1972, Bleem2015, Planck2016SZ, Hilton2018} and
optical/near-infrared \citep{Rykoff2016, WenHan2018, Gonzalez2019, Maturi2019}
surveys. These surveys, however, have yielded only a handful of confirmed
proto-cluster detections at $z\simgt 1.5$ \citep{Overzier2016}. The reason is
that, while at $z \simlt 1$--1.5 cluster cores are dominated by passive
early-type galaxies and are filled by hot gas, at higher $z$ cluster members
enter the dust-obscured star-formation phase and the intergalactic gas is no
longer necessarily at the virial temperature. In fact, several programs have
found an inversion of the star formation–density relation at $z\simgt 1.3$. At
lower redshifts, it has long been known that there is a well defined increase
of the passive elliptical and {lenticular (denoted S0)} population with
increasing density \citep{Dressler1980}. Dense cluster cores are preferentially
populated by massive, passively evolving, early-type galaxies. Star-forming
galaxies are generally found in the cluster outskirts and in the
field\footnote{{Field galaxies are those that do not reside in overdense
regions.}}. At higher redshifts, however, cluster cores are found to have an
increasing population of strongly star-forming, luminous infrared galaxies
\citep{Alberts2014, Alberts2016, Wagner2017}. Their specific SFR increases
rapidly from $z\sim 0.2$ to $z\sim 1.3$, mostly driven by the activation of
star formation in early-type galaxies. The star-formation activity in cluster
core ($r< 0.5\,$Mpc) galaxies reaches the field level at $z \simgt 1.2$.

\noindent So far, most of the detections of high-$z$ (proto-)clusters have been
obtained either as by-products of large spectroscopic or multi-band photometric
surveys, or using biased tracer techniques \citep{Overzier2016}. The latter
techniques consists in targeting the immediate environment of tracers of
massive forming systems, like high redshift radio galaxies and QSOs, Ly$\alpha$
blobs and bright sub-millimeter galaxies. The data sets collected in these ways
are obviously highly heterogeneous, affected by strong, hard to quantify
selection biases and only sparsely sample the redshift distribution of
(proto-)clusters. They are therefore unsuitable to obtain a complete picture of
the build-up of galaxy clusters over cosmic time.

\noindent The very fact that member galaxies are increasingly infrared luminous
with increasing redshift, means that (sub-)mm surveys are the most effective
tool to detect high-$z$ proto-clusters. They are rare objects and therefore the
multi-steradian CMB surveys are optimal to find at least the (sub-)mm brightest
ones. A blind search on \textit{Planck} maps was carried out by
\citet{PlanckXXXIX2016} at $5'$ resolution. They looked for intensity peaks
with ``cold'' sub-mm colours, i.e. with continuum spectra peaking between 353
and 857\,GHz, consistent with redshifts $z>2$ for typical dust emission
spectra. \textit{Herschel\/}/SPIRE follow-up of 234 \textit{Planck} targets
with such colours showed that almost all of them correspond to strong
over-densities of red 350 and $500\,\mu$m sources in comparison to reference
SPIRE fields \citep{PlanckHerschel2015}. Further investigations of
\textit{Planck} proto-cluster candidates were carried out by refs.
\citep{Clements2014, Clements2016, Greenslade2018}.

\noindent However, the \textit{Planck}'s angular resolution of $\simeq 5'$,
corresponding to a physical size of about 2.5\,Mpc at $z=1.5$--2, is not
optimal for detecting proto-clusters. By means of detailed simulations based on
a physically motivated galaxy evolution model, \citet{Negrello2017protocl}
showed that essentially all \textit{Planck}'s cold peaks can be interpreted as
positive Poisson fluctuations of the number of high-$z$ proto-clusters of dusty
galaxies within the \textit{Planck} beam, rather than being individual clumps
of physically bound galaxies.

\noindent The study \citep{Alberts2014} of 274 clusters with $0.3\le z \le 1.5$
from the \textit{Spitzer} InfraRed Array Camera (IRAC) Shallow Cluster Survey,
using \textit{Herschel}/SPIRE 250-$\mu$m imaging, showed that the density of
IR-emitting cluster members clearly exceeds that of the background field level
only within $0.5\,$Mpc of the cluster centre. A linear scale of 0.5\,Mpc
corresponds to an angular scale of about $1'$ at redshifts in the range
1.5--2.5, close to the PICO/CORE/CMB Bharat FWHM at 800\,GHz and to the
resolution of CMB-S4 and of the Simons Observatory at mm wavelengths. Thus,
next generation CMB experiments will be optimally suited to detect the bright
cluster cores of the kind discovered by \citet{Ivison2013} at $z=2.41$,
\citet{Wang2016} at $z=2.51$, \citet{Miller2018} at $z=4.31$ and
\citet{Oteo2018} at $z=4.0$ (see Fig.~\ref{fig:SEDprotocl}). Spectroscopic
confirmations of additional proto-clusters detected at sub-mm wavelengths have
been most recently reported by \citet{GomezGuijarro2019} and
\citet{Lacaille2018}. Figure~\ref{fig:SEDprotocl} shows that proto-clusters
bright enough to be detected by next generation CMB experiments exist out to
$z\simgt 4$. Note that, as argued by \citet{DeZotti2018}, proto-clusters stand
out as intensity peaks in background-subtracted maps more clearly than in
surveys of point sources. This is because such intensity peaks include the
contribution of all member galaxies, including those below the point source
detection limit, that may dominate the integrated flux density.

\noindent {How many dusty proto-cluster detections can we expect?} The
predictions of the model by \citet{Negrello2017protocl}, that reproduced all
the relevant data, are shown in Fig.~\ref{fig:protocl_counts}. Space-borne
experiments like PICO/CORE/CMB Bharat will detect many tens of thousands of
these objects, with a predicted redshift distribution peaking at $z\simeq 2$
and extending out to $z\simeq 4.5$.  Ground based instruments will
preferentially detect proto-clusters at higher redshifts, out to $z\simgt 5$,
with a distribution having a broad peak around $z\simeq 2.75$.   Since high-$z$
clusters are expected to be rare, the number of detections will be of only
$\simeq 10^{-2}\,\hbox{deg}^{-2}$ at the SPT completeness limit at 1.4\,mm. The
number of detections rapidly increase with decreasing detection limit, reaching
$\simeq 40\,\hbox{deg}^{-2}$ at the SPT confusion limit. This will constitute a
real breakthrough in the observational determination of the formation history
of the cluster-sized dark matter halos. Follow-up observations will
characterize the properties of member galaxies, probing the galaxy evolution in
dense environments and shedding light on the complex physical processes driving
it.

\begin{figure}
\begin{center}
\includegraphics[trim=0cm 0cm 0cm 0cm,clip=true, width=0.7\columnwidth]{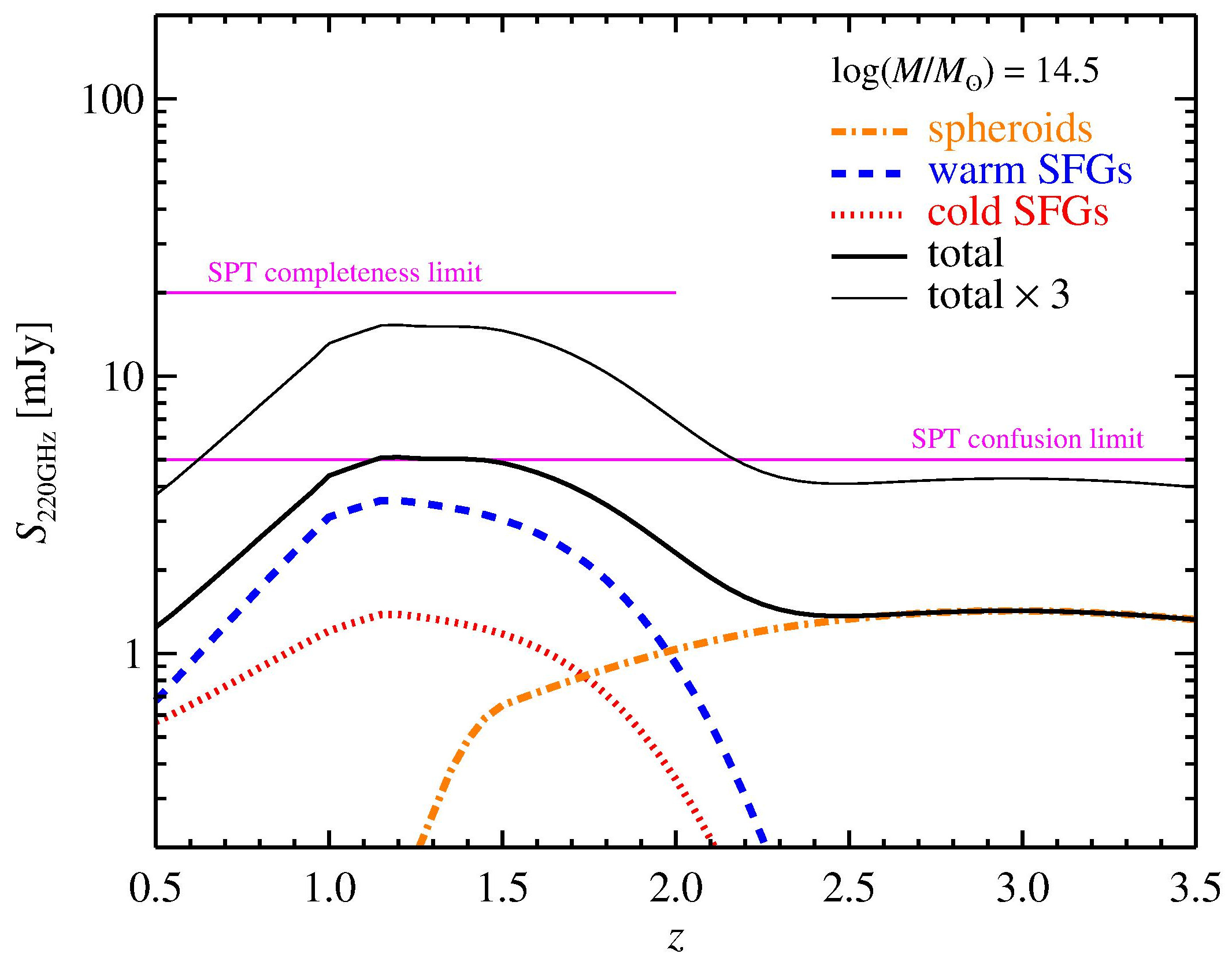}
\end{center}
\caption{Expected flux density at 220\,GHz due to the integrated dust emission from member galaxies
of a cluster with $M=10^{14.5}\,\hbox{M}_\odot$ as a
function of the cluster redshift (solid thick black curve). The cluster luminosity includes contributions
from normal late-type and starburst galaxies (warm and cold SFGs, respectively)
and from proto-spheroidal galaxies (spheroids), computed using the model by
\citet{Cai2013}. The thinner upper black
curve is a factor of 3 higher and illustrates the large variance of the cluster
IR emission at fixed mass \citep{Alberts2014, Alberts2016}. It must be noted
that the \textit{Planck} and the SPT surveys have detected $z>0.5$ clusters
with masses of more that $10^{15}\,\hbox{M}_\odot$. The upper horizontal line
corresponds to the SPT completeness limit, the lower one to its $5\,\sigma$ confusion limit. } \label{fig:IRcluster}
\end{figure}

\noindent CMB experiments will also allow us to investigate the evolution of
galaxy populations in clusters detected by other means, e.g. via their X-ray
emission or via the thermal SZ effect. \citet[][their Fig.~4]{DeZotti2018}
showed that space-borne CMB experiments with telescopes of the 1.5\,m class
will detect, at 800\,GHz the dust emission of the brightest $M \simeq
10^{14}\,\hbox{M}_\odot$ clusters and of typical $M \simeq
10^{14.5}\,\hbox{M}_\odot$ clusters at $1\simlt z \simlt 1.5$. More massive
clusters will be detected over broader redshift ranges.
Figure~\ref{fig:IRcluster} shows that SPT-like ground based surveys observing
at 220 GHz will perform similarly or only slightly worse. For comparison,
\textit{Herschel} has allowed the study of the IR emission from clusters up to
$z\simeq 1.7$ \citep{Alberts2016}, but the sample comprises only 11 clusters.
The \textit{Herschel} data have shown large variations in cluster properties,
highlighting the need for evolutionary studies of large, uniform cluster
samples over a broad redshift range. Next generation CMB experiments will
fulfill this need. Stacking will allow us to carry out a statistical
investigation of the evolution of the cluster IR emission to fainter levels.
Targets for stacking will abound. eROSITA (extended ROentgen Survey with an
Imaging Telescope Array) will provide an all-sky deep X-ray survey detecting
$\sim 10^5$ galaxy clusters out to $z>1$ \citep{Merloni2012}. CMB experiments
themselves will detect tens of thousands of clusters via the SZ effect
\citep{Hanany2019, Melin2018CORE, Abazajian2016, Ade2019}.

\noindent This will be important to investigate the evolution of the specific
SFR in dense environments. In addition, the dust emission in galaxy clusters
will impact the completeness of SZ surveys and will distort the SZ signal
\citep{Melin2018}, affecting the cosmological results derived from SZ
observations. While the latter effect was shown to be negligible in
\textit{Planck}'s case \citep{Melin2018}, it will be important for the much
deeper next generation surveys.

\begin{figure}
\begin{center}
\includegraphics[width=\columnwidth]{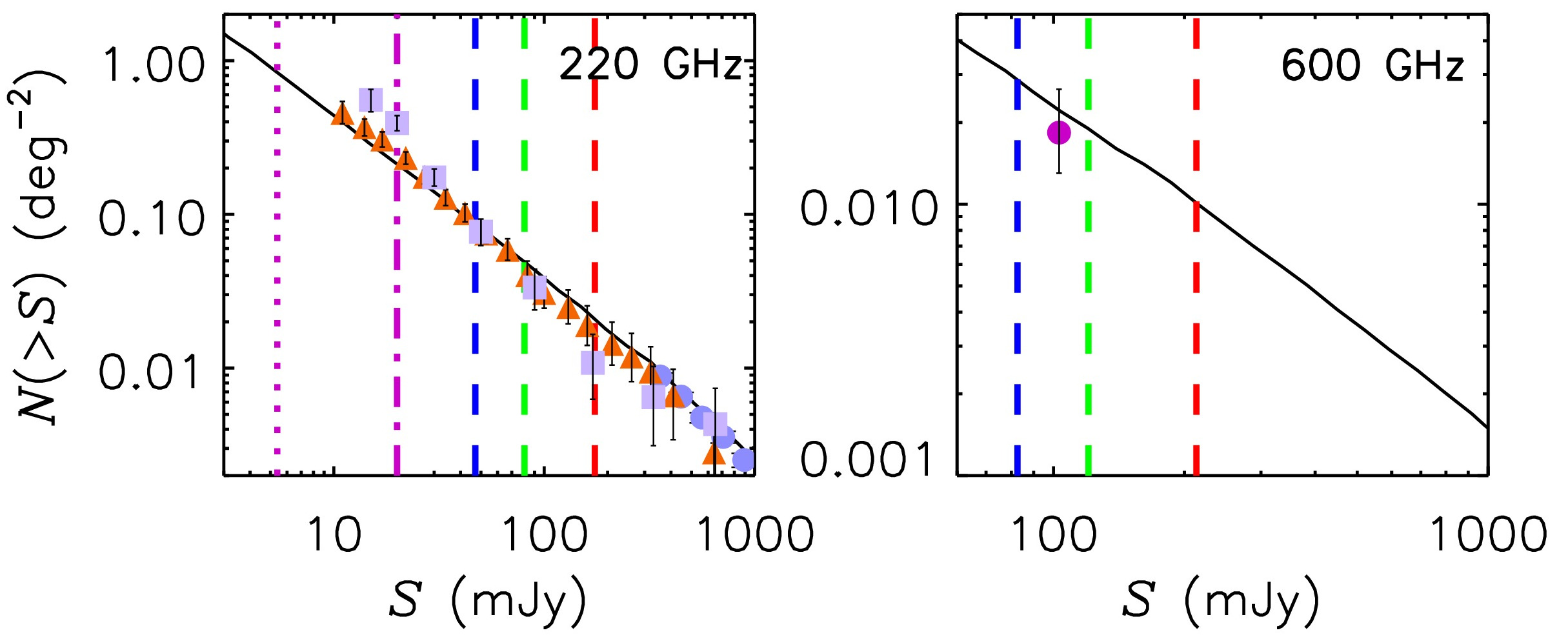}
\end{center}
\caption{Integral number counts of radio sources at 220 and 600 GHz. The vertical dashed
lines show the $5\,\sigma$ confusion limits for a space-borne experiment with a 1\,m, 1.5\,m and a 2\,m telescope
(from right to left) operating at the diffraction limit. The dot-dashed and dotted vertical lines on the
left panel show the completeness limit of the SPT survey and the $5\,\sigma$ confusion limit for the SPT
telescope, respectively. The data points on the left panel are from \citet[][SPT, orange triangles)]{Mocanu2013},
\citet[][ACT, lavender squares]{Marsden2014}, \citet[][light blue circles)]{PlanckCollaborationXIII2011};
the data point on the right panel is from \citet{Bonato2019a} and is based on \textit{Herschel} survey data.
The solid black lines are predictions of the \citet{Tucci2011} model. }\label{fig:radio_counts}
\end{figure}

\section{Radio sources}\label{sect:radio}

\subsection{Blazar physics}

\noindent Although substantial progress on the characterization of mm and
sub-mm properties of extragalactic radio sources has been made in recent years
mainly thanks to surveys with  WMAP, \textit{Planck}, the SPT and the ACT, the
available information is still limited. The overwhelming majority of
extragalactic radio sources detectable in the frequency range of CMB
experiments are blazars,  i.e. sources whose radio emission is dominated by
relativistic jets collimated by intense magnetic fields and closely aligned
with the line of sight. These objects with extreme properties are of special
interest since they are also strong gamma-ray sources:  about 90\% of the
firmly identified extragalactic  gamma-ray sources are blazars.

\noindent Accurate source counts over large flux density intervals provide key
constraints  on evolutionary models of these sources. Just because high
frequency surveys are still far less extensive than those at low radio
frequencies, evolutionary models for blazar populations, Flat Spectrum Radio
Quasars (FSRQs) and BL Lacertae sources (BL Lacs), are far less advanced than
those for steep-spectrum radio sources. For example, while clear evidence for
downsizing\footnote{``Downsizing'' refers to the very different evolutionary
behaviour of high- and low-luminosity sources, in the sense that the redshift
of the peak space density of sources decreases with luminosity.} was reported
in the case of steep-spectrum sources \citep{Massardi2010, Rigby2015}, the
available data are insufficient to test if this is the case also for FSRQs; for
BL Lacs the constraints on evolutionary parameters are even weaker. This
situation hampers sharp tests of unified models.

\noindent \textit{Planck} has already provided strong indications of the
crucial role of blazar photometry up to sub-mm wavelengths to get information
on the energy spectrum of relativistic electrons responsible for the
synchrotron emission, with interesting implications for the acceleration
mechanisms \citep{PlanckCollaborationXLV2016}.

\noindent Another interesting open question is the geometry of the emitting
regions. The most commonly used model for the spectral energy distribution
(SED) of compact, radio loud Active Galactic Nuclei (AGNs) is a leptonic,
one-zone model, where the emission originates in a single component. The SEDs
typically consist of two broad-band bumps; the one at lower frequencies is
attributed to synchrotron radiation while the second, peaking at gamma-ray
energies, is attributed to inverse Compton. The one-zone model is generally
found to provide an adequate approximation primarily because of the limited
observational characterization of the synchrotron SED, with fragmentary data
over a limited frequency range. However VLBI images show multiple knots often
called ``components'' of the jet. The standard model \citep{MarscherGear1985}
interprets the knots as due to shocks that enhance the local synchrotron
emission.

\noindent The spectrum is explained as the result of the superposition of
different synchrotron self-absorbed components in a conical geometry. The
synchrotron self-absorption optical depth scales as $\tau_{\rm sync} \propto
B_\perp^{(p+2)/2} \nu^{-(p+4)/2}$ where $B_\perp$ is the magnetic field
component perpendicular to the electron velocity and $p$ is spectral index of
the energy distribution of relativistic electrons (typically, $p \simeq 2.5$).
Thus $\tau_{\rm sync}$ increases towards the nucleus as the magnetic field
intensity and its ordering increases, but is strongly frequency dependent: the
emission at higher and higher frequencies comes from smaller and smaller
distances from the central engine.

\noindent Thus the mm and sub-mm emissions provide information on the innermost
regions of the jets, where it is optically thin, while the emission at longer
wavelengths is affected by self-absorption. Interestingly millimeter-wave flux
densities of FSRQs turn out to be strongly correlated with simultaneous
gamma-ray fluxes \citep{Fuhrmann2016, FanWu2018}. The strongest gamma-ray
flares were found to occur during the rising/peaking stages of millimeter
flares. This suggests that the gamma-ray flares originate in the
millimeter-wave emitting regions of these sources.

\noindent The available data are mostly at cm (or longer) wavelengths and are
scanty at (sub-)mm wavelengths because of the limited sky areas covered by the
available surveys. Ground-based experiments with $\sim 6\,$m telescopes, like
CMB-S4 and the Simons Observatory,  will detect thousands of blazars per sr at
millimeter wavelengths (Fig.~\ref{fig:radio_counts}). Extrapolating the
8.4\,GHz flux densities of the 18 $z>4$ FSRQs listed by \citet{Caccianiga2019}
using the measured 1.4--8.4\,GHz spectral indices we find that a large fraction
of them will be detected by these experiments, including the highest redshift
blazar known, GB6J$090631+693027$ at $z=5.47$. The radio selection, being
unaffected by obscuration, provides an unbiased census at least of the
radio-loud fraction of high-$z$ AGNs.

\noindent The most luminous high-$z$ FSRQs were found to have black holes with
the largest masses, up to $\simeq 4\times 10^{10}\,M_\odot$
\citep[S5\,$0014+813$ at $z=3.366$;][]{Ghisellini2013}. Such objects have
particularly hard mm-wave spectra, are rare and bright because of the Doppler
boosting of their flux densities. CMB surveys are thus optimally suited to
detect them. Since the flux boosting occurs for jets closely aligned with the
line of sight ($\theta < 1/\Gamma$, $\Gamma \sim 15$ being the bulk Lorentz
factor), for each FSRQ there are other $2 \Gamma^2$ (i.e. hundreds) sources of
similar intrinsic properties but pointing elsewhere. This means that blazars
are very efficient probes of extreme super-massive black holes at high $z$.

\noindent Very large black hole masses at high $z$ are puzzling because it is
challenging to grow a stellar mass seed black hole to $> 10^9\,M_\odot$ in the
limited age of the universe. It is even more challenging in the case of jetted
quasars because it is commonly believed that the jets are associated with
rapidly spinning black holes. But then the radiative efficiency is large and
the mass growth is slower. Yet at least 4 FSRQs has been discovered at $z>5$;
one of them (SDSS\,J$013127.34–032100.1$ at $z = 5.18$) has estimated black
hole mass of $\simeq 1\times 10^{10}\,M_\odot$ \citep{Ghisellini2015}.

\noindent Next-generation space-borne CMB experiments with $\simeq 1.5\,$m
telescopes, like PICO, will provide, for the first time, samples of hundreds of
blazars blindly selected at sub-mm wavelengths. An important property of
surveys from space is that they provide simultaneous photometry over a broad
frequency range, thus overcoming the complications due to variability and
allowing us to directly connect the observed SED to the physical processes
operating along the jet.

\subsection{Earliest and latest phases of radio activity}

\noindent Large-area surveys at  frequencies of tens to hundreds GHz will also
detect the rare but very interesting sources associated to the earliest and to
the latest stages of the radio-AGN evolution, both characterized by emissions
peaking in this frequency range \citep{DeZotti2005}. It is now widely agreed
that extreme  gigahertz peaked spectrum (GPS) sources correspond to the early
stages of the evolution of powerful radio sources, when the radio emitting
region grows and expands within the interstellar medium of the host galaxy,
before plunging in the inter-galactic medium and becoming an extended radio
source. There is a clear anti-correlation between the peak (turnover) frequency
and the projected linear size of GPS sources, suggesting a decrease of the peak
frequency as the emitting blob expands. The identification of these sources is
therefore a key element in the study of the early evolution of radio AGNs.
High-frequency surveys will detect these sources very close to the moment when
they turn on.

\noindent Possible examples of extremely young sources are the six narrow-line
Seyfert 1 galaxies detected by \citet{Lahteenmaki2018} at 37 GHz with flux
densities of 270--970\,mJy but undetected by the FIRST survey, complete down to
$\simeq 1$\,mJy at 1.4\,GHz, carried out about 20\,yr ago. One possibility is
that the new observations have discovered newly triggered radio activity from
nuclei that were essentially radio silent two decades ago.

\noindent Young radio activity was recently discovered by \citet{Bruni2019} in
the nuclei of 8 out of 13 ($\simeq 61\%$) hard X-ray selected giant radio
galaxies for which they had enough spectral coverage to ascertain the presence
of a peak. Two of the 8 sources have a peak frequency $>10\,$GHz and at least
one is bright enough to be clearly detected by next generation CMB experiments.

\noindent The multi-frequency surveys by next generation CMB experiments will
provide an unbiased view of the frequency of these phenomena and will measure
their high frequency SEDs, shedding light on their nature. These data will
enable studies of the launching of relativistic jets as well as of the
evolutionary paths that young AGNs take on their way to becoming fully-evolved,
powerful radio sources.

\noindent Large area (sub-)mm surveys will also allow us to investigate the
late stages of the  AGN evolution in elliptical galaxies, characterized by low
radiation/accretion efficiency. These manifest themselves via a nuclear radio
emission described by advection-dominated accretion flows (ADAFs) and/or by
adiabatic inflow-outflow solutions (ADIOS).  \citet{Doi2005} have found that at
least half of their sample of 20 low-luminosity AGNs with compact radio cores
show radio spectra rising at least up to 96 GHz, consistent with the
`sub-millimetre bump’ predicted by an ADAF model. Again CMB surveys will
determine the abundance of these objects and their SED measurements will
clarify the origin of the emission.

\noindent Predictions of the expected number of detections of early and late
phases of radio activity are limited to $\le 30\,$GHz \citep{DeZotti2005,
TintiDeZotti2006}. They suggest that at 20-30 GHz hundreds of these objects can
be detected by ground-based experiments. Space-borne experiments are
confusion-limited to much brighter flux densities, implying a detection rate at
least one order of magnitude lower. The number of detections is predicted to
drop rapidly with increasing frequency \citep{BlandfordMcKee1976,
GranotSari2002}.

\subsection{The extragalactic transient sky}

\noindent High-sensitivity and high-angular-resolution CMB surveys also offer a
unique opportunity to carry out an unbiased investigation of the largely
unexplored extragalactic mm/sub-mm transient sky \citep{Metzger2015}. So far
measurements have been limited to follow-up of objects detected at other
wavelengths, with limited success partly because of the need for excellent
weather conditions or because the events were too short-lived to detect without
very rapid reaction times. CMB surveys will allow us to discover new, unknown
transient sources in this band.

\noindent One example of transient phenomena are outbursts from AGNs and
especially from blazars. Outbursts and, more generally, variability, provide
key information on the flow of the plasma within the relativistic jets.
Signatures of evolving shocks in the strongest radio flares were seen by
\citet{PlanckCollaborationXLV2016} although {high frequency light curves are
generally quite similar (approximately achromatic variability)}. These results
are compatible with the standard shocked jet model, but other interpretations
are possible. Definite conclusions are currently hampered by the limited
statistics. This limit will be overcome by next generation CMB experiments from
space which will provide multi-epoch simultaneous observations of  large blazar
samples over a broad frequency range. This will allow us to study their
variability properties as a function of flux density and spectral shape.

\noindent Perhaps even more interesting is the possibility of detecting radio
aftergrows of gamma-ray bursts (GRBs). Afterglows often have a spectral peak in
or near the mm range \citep{GranotSari2002}, with emission lasting over days to
weeks. One candidate object with properties broadly consistent with a GRB
afterglow was tentatively detected by \citet{Whitehorn2016} on SPT data over
$100\,\hbox{deg}^2$ with an observing time of 6,000\,h, but the statistical
significance of the detection was too low to completely rule out a fluctuation.
GRB emission is expected to be less tightly beamed at these wavelengths than in
gamma-rays. Thus afterglows not accompanied by detectable gamma-ray emission
are expected to exist, but have not been detected yet. Blind surveys of large
sky areas by next generation ground-based CMB experiments down to $\simeq
10\,$mJy sensitivity  can reveal these orphan afterglows and new, unknown
sources. Multiple detections per year are expected. Even a non-detection will
place interesting constraints on the shock dynamics and on the energy budget of
the unknown GRB progenitors.

\noindent One example of unexpected phenomena that may show up at (sub-)mm
wavelengths is the extraordinary extragalactic transient AT2018cow, with an
estimated peak flux density of 94 mJy at $\simeq 100\,$GHz \citep{Ho2019}. This
object may herald a new class of energetic transients which at early times are
most readily observed at (sub-)mm wavelengths.

\begin{figure}
\begin{center}
\includegraphics[width=0.8\columnwidth]{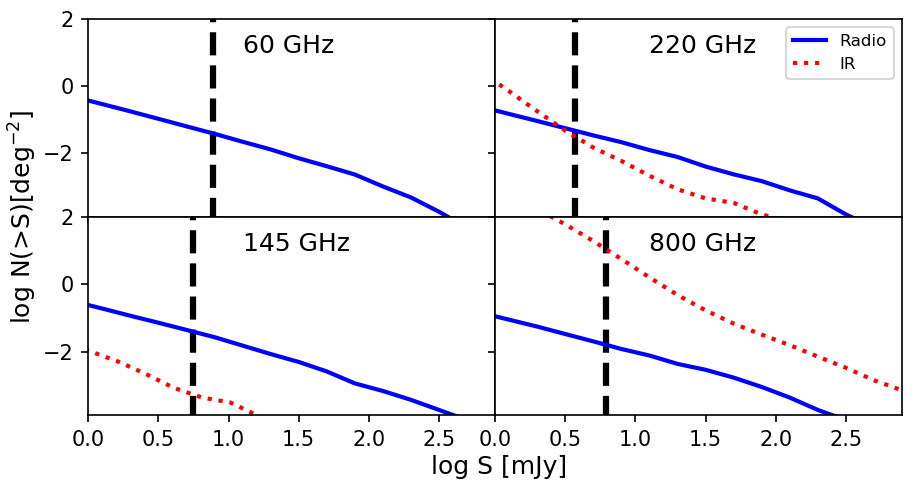}
\caption{Estimated integral number counts in polarized intensity of radio sources and
of dusty galaxies (IR) at {60, 145, 220 and 800\,GHz. The contribution of dusty galaxies is completely
negligible at 60\,GHz.} The vertical dashed black lines show
the $5\,\sigma$ detection limits for a space-borne instrument with a 1.5\,m telescope
and state-of-the-art sensitivity derived from simulations similar to those described in \citet{DeZotti2018}.
The present simulations assumed log-normal distributions of the polarization fractions with mean
and dispersion of 2.14\% and of 0.9\%, respectively, for radio sources and of 1.4\ and 1\%,
respectively, for dusty galaxies.}
\label{fig:pol_counts}
\end{center}
\end{figure}


\section{Detecting sources in polarization}\label{sect:counts_pol}

\noindent Accurate simulations \citep{Remazeilles2018} showed that, for a
tensor-to-scalar ratio $r\simeq 10^{-3}$ (i.e. at levels predicted by models
currently of special interest, such as Starobinsky's $R^2$ and Higgs
inflation), the overall uncertainty on $r$ is dominated by foreground residuals
and that unresolved polarized point sources can be the dominant foreground
contamination over a broad range of angular scales ($\ell \simgt 50$). A
thorough understanding of the polarization properties of extragalactic sources
is therefore crucial.



\noindent While the point source power spectra in total intensity are quite
well constrained, estimates in polarization are obtained {by} coupling the
counts in total intensity with distributions of the polarization fractions
derived from lower frequency surveys. This implicitly assumes that such
distributions are frequency-independent. Although this assumption is consistent
with the available data \citep{Battye2011, Galluzzi2018}, there are substantial
uncertainties and indeed variations are expected since emissions at different
frequencies of flat-spectrum sources, that dominate in the relevant frequency
range, come from different regions where the magnetic field properties are
expected to be different.

\noindent Polarization measurements at $\simgt 100\,$GHz are still scanty. Only
a few tens of sources (all radio) were detected by \textit{Planck} in the
`extragalactic zone' \citep{DeZotti2018}. \citet{Galluzzi2019} carried out ALMA
high sensitivity polarimetric observations at 97.5 GHz of a complete sample of
32 compact extragalactic radio sources brighter than 200 mJy at 20 GHz; a
detection rate of 97\% was achieved.

\noindent \citet{Datta2019} reported the detection of linear polarization at
148\,GHz of 26 extragalactic sources with total flux density $S_{148\,\rm GHz}>
215\,$mJy, 14 of which at greater than $3\,\sigma$ significance, during the
first two seasons of the ACT polarization (ACTPol) survey, covering
$680\,\hbox{deg}^2$ of the sky. Their results are consistent with a mean
fractional polarization $\langle \Pi\rangle = 0.028\pm 0.005$ and a standard
deviation $\sigma_{\rm p} = 0.054$, independent of total intensity.

\noindent \citet{Gupta2019} investigated the polarisation properties of
extragalactic sources in the SPTpol $500\,\hbox{deg}^2$ survey at 95 and
150\,GHz down to 6\,mJy. They found no evidence that the polarisation fraction
depends on flux density. Assuming $\langle \Pi\rangle$ to be constant across
flux bins they obtained, for radio sources, flux weighted $\langle \Pi^2
\rangle =(8.9\pm 1.1)\times 10^{-4}$ at 95\,GHz and $\langle \Pi^2 \rangle
=(6.9\pm 1.1)\times 10^{-4}$ at 150\,GHz.

\noindent These results are consistent with the conclusion by
\citet{Bonavera2017a} who applied stacking techniques to radio sources detected
by \textit{Planck} in total intensity, finding an average fractional
polarization of $\simeq 3\%$, essentially independent of frequency from 30 to
353\,GHz. A similar conclusion was reached by the independent analysis of
\citet{Trombetti2018}.

\noindent The (sub-)mm polarization properties of dusty galaxies are
essentially unknown. The only published measurement of the polarization degree
averaged over the whole galaxy \citep{GreavesHolland2002} yielded $\Pi\simeq
0.4\%$ at $850\,\mu$m for the prototype starburst galaxy M\,82. A 99\%
confidence upper limit of 1.54\%  at the same wavelength was reported by
\citet{Seiffert2007} for the ultraluminous infrared galaxy (ULIRG) Arp\,220.
From the \textit{Planck} dust polarization maps of the Milky Way,
\citet{DeZotti2018} found an average value of the Stokes $Q$ parameter of about
2.7\%. If this is typical of disk galaxies, their mean polarization degree
averaged over all possible inclination angles is  $\simeq 1.4\%$.

\noindent By applying the stacking techniques to a large sample of dusty
galaxies drawn from the PCCS2 857 GHz catalogue, \citet{Bonavera2017b}
estimated a median $\Pi$ of ($2.0 \pm 0.8\%$) at 353 GHz, consistent with the
90\% confidence upper limit of 2.2\% derived by \citet{Trombetti2018}. The
sample by \citet{Gupta2019} includes 55 sources classified as dusty galaxies.
No polarization signal was detected for these sources. The resulting 95\%
confidence level upper limits are $\langle \Pi^2 \rangle < 16.9 \times 10^{-3}$
at 95\,GHz and $\langle \Pi^2 \rangle < 2.6 \times 10^{-3}$ at 150\,GHz,
consistent with earlier results.

These low values of $\Pi$ are understood as due to the complex structure of
galactic magnetic fields, with reversals along the line of sight, and to the
disordered alignment of dust grains; both effects work to decrease the global
polarization degree. Nevertheless, {the current limits on the distribution of
polarization fractions of dusty galaxies permit a contamination of CMB
polarization maps comparable to that of radio sources down to 100--140\,GHz,
and dominant at higher frequencies \citep{Bonavera2017b, Trombetti2018}. The
amplitude of the power spectrum of polarized dusty galaxies} may be close to
the level of CMB lensing $B$-modes and of primordial $B$-modes for $r\simeq
0.01$.

\noindent At variance with total intensity, in the case of polarization the
detection limit is dictated by sensitivity, not by confusion noise. Hence, the
spectacular improvement in sensitivity of next generation CMB experiments,
compared to \textit{Planck}, will allow a real breakthrough in the
characterization of the polarization properties of extragalactic sources.
Figure~\ref{fig:pol_counts} shows the integral number counts in polarized
intensity of radio sources and of dusty galaxies at 60, 145, 220 and 800\,GHz
obtained from the simulations described in \citet{DeZotti2018}. The vertical
dashed black lines show the $5\,\sigma$ detection limits for a space-borne
instrument with a 1.5\,m telescope and state-of-the-art sensitivity, derived
from simulations made as described in \citet{DeZotti2018}, assuming log-normal
distributions of the polarization fractions. For radio sources we adopted the
same mean and dispersion (2.14\% and of 0.9\%, respectively) used by
\citet{DeZotti2018}. For dusty galaxies we adopted the value for our own
Galaxy, averaged over inclinations (1.4\%), with a dispersion of 1\%. The
estimated $5\,\sigma$ detection limits are 3.7\,mJy at 220\,GHz and 6.2\,mJy at
800\,GHz. At these limits the expected integral counts at 220\,GHz  are of
$\simeq 160\,\hbox{sr}^{-1}$ radio sources and of $\simeq 100\,\hbox{sr}^{-1}$
dusty galaxies; at 800\,GHz they are of $\simeq 50\,\hbox{sr}^{-1}$ radio
sources and of $\simeq 36,000\,\hbox{sr}^{-1}$ dusty galaxies.

{Under these assumptions, radio sources are the dominant population below
$\simeq 200\,$GHz. } A comparison with the results reported by
\citet{DeZotti2018} who used a mean polarization fraction of 0.5\% for dusty
galaxies shows that the predicted counts are highly sensitive to the choice for
this, highly uncertain, quantity.

\noindent As illustrated by Fig.~\ref{fig:pol_counts}, next generation CMB
experiments will be capable of providing, for the first time, direct counts in
polarization both for radio sources and for dusty galaxies, thus overcoming the
current large uncertainties on the source power spectra in polarization. On one
side this will allow a much better control of the extragalactic source
contamination of CMB maps. This is particularly important  in the 60--120\,GHz
frequency range, where  diffuse polarized foreground emissions display a broad
minimum.

\noindent On the other side, polarization observations enable us to understand
geometrical structure and intensity of magnetic fields, particle densities and
structures of emission regions. In the case of extragalactic radio sources,
emission at mm/sub-mm wavelengths is synchrotron radiation arising close to the
origin of the jet, on sub-parsec scales, generally unresolved even by the
highest frequency very long baseline interferometry (VLBI) maps
\citep{Nartallo1998}. At these wavelengths the emission is expected to be
optically thin, so that self-absorption and Faraday rotation are negligible.
Then, in principle, the linear polarization degree can be as high as 60--80\%
if the magnetic field is ordered \citep{SaikiaSalter1988}. Hence, measurements
of the polarization degree constrain the magnetic field geometry.

\section{Conclusions}\label{sect:conclusions}

\noindent Thanks to their high sensitivity and to the coverage of large
fractions of the sky, next generation CMB experiments will provide
ground-breaking results in the field of extragalactic astrophysics. They will
provide samples of several thousands of the brightest high-$z$ strongly lensed,
dusty galaxies, with extreme amplifications, up to $\mu \simeq 50$, and out to
$z\simgt 6$. This will constitute an ideal data-base for high-resolution
follow-up observations addressing the internal structure and kinematics of
primordial galaxies, i.e. to understand the physical processes that drive the
galaxy formation and early evolution across a broad redshift range, up to the
re-ionization epoch.

\noindent CMB experiments will also provide unbiased, flux limited samples of
tens of thousands of dense proto-cluster cores out to $z\simgt 4$, well beyond
the reach of classical (optical, X-ray, SZ) cluster surveys that are mostly
limited to $z\simlt 1.5$. They will effectively open a new window on the study
of early phases of cluster formation, when their member galaxies were actively
star forming and before the hot intergalactic medium was in place. This is
crucial to observationally assess the formation history of the most massive
dark matter halos, traced by clusters, a critical test of models for structure
formation.

\noindent These experiments will also allow us to investigate, via direct
detections complemented with stacking analyses, the evolution of the
star-formation rate in virialized galaxy clusters detected by X-ray and SZ
surveys (including those carried out by the experiments themselves), shedding
light on the role of dense environments on galaxy evolution.

\noindent The data on radio sources will greatly improve our understanding of
the evolutionary properties of FSRQs and BL Lac objects. They will also probe
the jet physics in its innermost regions as well as the earliest and latest
phases of radio activity. Particularly interesting is the possibility of
getting an unbiased view of the abundance of candidate newly-born radio
sources. These surveys also offer a unique opportunity to carry out an unbiased
investigation of the largely unexplored mm/sub-mm transient sky, including the
detection of predicted, but still unseen, ``orphan'' radio afterglows of GRBs
as well as unexpected transient phenomena.

\noindent Furthermore, these experiments will provide the first, extensive,
blind high-frequency census of the polarization properties of radio sources and
of star-forming galaxies. This is essential to clean CMB maps at the level
required to measure the faint primordial $B$-mode power spectrum. At the same
time, mm/sub-mm polarization data on radio sources provide unique information
on the geometry of the magnetic field on sub-pc scales, unresolved even by
high-frequency VLBI observations. Polarimetry of dusty galaxies is informative
on the structure of galactic magnetic fields.

\section*{Conflict of Interest Statement}

The authors declare that  the research was conducted in the absence of any
commercial or financial relationships that could be construed as a potential
conflict of interest.

\section*{Author Contributions}

GDZ has coordinated the work and written most of the text. MB, MN and ZYC have
made model calculations and prepared Figs.\,2, 3, 5, 6, 7, 8 and 9. TT and CB
worked on the selection of strongly lensed galaxies detected by \textit{Planck}
and made Fig.\,4. DH and MLC made the simulations and the source extraction on
degraded SPT maps and produced Fig.\,1. ZYC computed the redshift distribution
of strongly lensed galaxies at 1.4\,mm and the integrated flux density of
cluster galaxies at 220\,GHz as a function of redshift, and made a panel of
Fig.\,3 and Fig.\,8. LL and JGN performed the simulations used to determine the
detection limits in polarization and made Fig.\,10. All authors critically
reviewed the entire paper.

\section*{Funding}
GDZ, CB and TT acknowledge financial support from ASI/INAF agreement
n.~2014-024-R.1 for the {\it Planck} LFI Activity of Phase E2 and from the
ASI/Physics Department of the university of Roma--Tor Vergata agreement n.
2016-24-H.0 for study activities of the Italian cosmology community. MB
acknowledges support from the Italian Ministero dell'Istruzione, Universit\`a e
Ricerca through the grant `Progetti Premiali 2012-iALMA' (CUP C52I13000140001)
and, together with CB and TT, from INAF under PRIN SKA/CTA FORECaST. DH
acknowledges partial financial support from the Spanish Ministerio de
Econom{\'\i}a y Competitividad (MINECO) project AYA2015- 64508-P and from the
RADIOFOREGROUNDS project, funded by the European Comission's H2020 Research
Infrastructures under the Grant Agreement 687312.

\section*{Acknowledgments}
{We are grateful to the referees for a careful reading of the manuscript and
many useful comments.} GDZ acknowledges enlightening discussions on CMB
experiments with S. Hanany and J. Delabrouille.



\bibliographystyle{frontiersinHLTH&FPHY} 
\bibliography{frontiers}


\end{document}